\documentclass[a4paper,10pt]{article}
\usepackage{graphicx}
\usepackage{amssymb}

\begin{document}

\begin{center}

{\LARGE 
{\bf Note on the inelastic neutron scattering spectrum 
in cuprate superconductors}
}

\vspace{5mm}

O. Narikiyo

{\sl Department of Physics, Kyushu University, Fukuoka 819-0395, Japan}

\vspace{3mm}

(August 29, 2016) 

\vspace{5mm}

\end{center}

\begin{abstract}
The inelastic neutron scattering spectrum 
in cuprate superconductors 
is discussed on the basis of 
the itinerant-localized duality model for strongly correlated electrons. 
In Appendix 
the consistency with recent rigorous theoretical result on ARPES is discussed. 


\end{abstract}

\vspace{10mm}

New experimental findings have been continuously brought 
by inelastic neutron scattering 
in cuprate superconductors. 
In recent years the spin excitation at higher energies 
than the resonance energy has been intensively discussed\cite{Levi}. 

For optimally doped La$_2$SrCuO$_4$ 
the spectral intensity in ${\bf q}$-$\omega$ space 
has been established experimentally\cite{LSCO}. 
In the following 
we give a phenomenological description 
of this completely established spectrum 
on the basis of the itinerant-localized duality model\cite{MN}. 

In the study of cuprate superconductors 
the duality model is effective to understand 
\lq\lq pseudogap" and \lq\lq spin-charge separation" anomalies\cite{MN} 
in the normal phase. 
Especially the \lq\lq pseudogap" phenomena 
are understood as a spectral-weight transfer\cite{MN,NM1,NM2,Narikiyo,NM3} 
from itinerant to localized degrees of freedom. 
Such a spectral weight transfer is missed 
in the standard spin-fluctuation theory\cite{Moriya} 
where the magnetically ordered phase is metallic. 
Since the ordered phase is the Mott insulator 
in the case of cuprates, 
the itinerant spectral weight 
representing the metallic component 
should vanish 
approaching the Mott transition 
and the spectral weight transfer 
from low-energy itinerant 
to high-energy localized spectral weights 
is inevitable. 

Above mentioned localized part 
representing the sloppy spin-wave\cite{Narikiyo,NM3} is 
the spin excitation at higher energies 
than the resonance energy\cite{Levi,LSCO}. 
This localized part is converted to the spin-wave spectrum 
in the Mott insulator. 

The itinerant-localized crossover\cite{NM1,NM2,NM4} 
resulting from the spectral weight transfer 
is observed experimentally\cite{phase-diagram}. 
The pseudogap at lowest temperature or lowest energy 
should be ascribed to the superconducting fluctuations\cite{phase-diagram}. 

\vskip 30pt

\noindent
{\Large\bf Appendix}

\vskip 20pt

\noindent
{\large\bf Summary of Appendix}
\vskip 10pt

In his recent review article 
Kohno has summarized a unified understanding 
of the single-particle excitation spectrum near the Mott transition. 
It seems that his work gives a rigorous support 
to previous studies based on the sloppy spin-wave in the metallic state. 

\vskip 20pt
\noindent
{\large\bf Introduction}
\vskip 10pt

In his recent review article~\cite{Kohno} 
Kohno gives a detailed explanation of the single-particle excitation spectrum 
near the Mott transition in a unified manner 
based on the exact results for the Hubbard model. 
Although he explains various aspects of the spectrum, 
I think that the most important finding is 
the existence of the spectrum related to the spin-wave in the insulating state. 

\vskip 20pt
\noindent
{\large\bf Primary: spin excitation}
\vskip 10pt

In strongly correlated electron systems 
the charge excitation is suppressed 
so that the spin excitation becomes dominant in the low-energy region. 
The salient feature of the spin-excitation spectrum near the Mott transition 
is the presence of the sloppy spin-wave. 
This spin-wave is described by the localized component 
of the itinerant-localized duality model. 
The spectral weight is transferred 
from the itinerant component to the localized component 
approaching the Mott transition. 
This spectral-weight transfer is observed 
as the pseudo-gap of the spin-excitation spectrum. 
At the Mott transition 
the itinerant component vanishes so that only the localized component remains. 
The elementary excitation of the remaining localized component is the spin-wave. 
The remnant of the spin-wave survives 
in the metallic state near the Mott insulator. 
This point is confirmed both theoretically and experimentally 
as discussed in the main text of this note. 

\vskip 20pt
\noindent
{\large\bf Secondary: single-particle excitation}
\vskip 10pt

Once the spin-excitation spectrum is established as mentioned above, 
the calculation of the single-particle excitation spectrum is the secondary task. 
Such a task is performed by many authors~\cite{KS}. 
The spectral weight of the sloppy spin-wave in $({\bf q},\omega)$-space 
is reflected in the single-particle spectral weight in $({\bf k},\varepsilon)$-space 
via the coupling between these two excitations. 
It seems that Kohno's work is consistent with 
and gives a rigorous support to previous developments. 

\vskip 20pt
\noindent
{\large\bf Conclusion of Appendix}
\vskip 10pt

Consequently we can conclude 
that the sloppy spin-wave in the metallic state 
is the key ingredient to understand the Mott transition.

\end{document}